\documentclass[sigconf,nonacm]{aamas} 

\usepackage{balance} % for balancing columns on the final page

\title[Shortlisting: a Principled Approach]{Shortlisting: a Principled Approach}

\author{Edith Elkind}
\affiliation{
  \institution{Northwestern University}
  \city{Evanston}
  \country{United States}}
\email{edith.elkind@northwestern.edu}

\author{Qishen Han}
\affiliation{
  \institution{Rutgers University}
  \city{New Brunswick}
  \country{United States}}
\email{hnickc2017@gmail.com}

\author{Lirong Xia}
\affiliation{
  \institution{Rutgers University}
  \city{New Brunswick}
  \country{United States}}
\email{lirong.xia@rutgers.edu}

\begin{abstract}
Shortlisting is the process of selecting a subset of alternatives from a larger pool for further consideration or final decision-making. It is widely applied in social choice and multi-agent system scenarios. The growing demand for participatory decision-making and the continuously expanding space of candidates create an urgent need for efficient and fair shortlisting procedures. However, little principled study has been done on this problem. This blue-sky paper aims to highlight the overlooked significance of shortlisting, distinguish it from related problems, provide initial thoughts, and, more importantly, serve as a call to arms. We envision that principled shortlisting can reduce cognitive burden, enable fair collective decisions, encourage broader participation, and ultimately build trust in democratic systems.
\end{abstract}


\newcommand{\BibTeX}{\rm B\kern-.05em{\sc i\kern-.025em b}\kern-.08em\TeX}

\usepackage{diagbox}

\usepackage{booktabs} % For formal tables
\usepackage{algorithm}
\usepackage{algorithmic}
\usepackage{pifont}
\usepackage{nicefrac}
\usepackage{color}
\usepackage{xcolor,colortbl}
\usepackage{wrapfig}
\usepackage{graphicx}
\usepackage{mathtools}
 
\usepackage{bm}
\usepackage{rotating}
\usepackage{multirow} 
\usepackage{enumitem}
\usepackage{hyperref}[backref=page]
\hypersetup{
     colorlinks   = true,
     linkcolor    = red, % color of internal links
     urlcolor     = blue, % color of external links
	 citecolor    = blue % color of links to bibliography
}

\newtheorem{dfn}{Definition}

\newtheorem{ex}{Example}

\newcounter{newct}

\newcommand{\bv}{\begin{array}}

\newcommand{\ma}{\mathcal A}
\newcommand{\prefspace}{\mathcal E}

\newcommand{\mL}{\mathcal L}

\newcommand{\ra}{\rightarrow}

\newcommand{\shortlist}{{\mathcal S}}

\newcommand{\myparagraph}[1]{\vspace{1mm}\noindent {\bf\boldmath #1}}

\newcommand{\Omit}[1]{}

\newcommand{\decspace}{{\mathcal D}}

\newcommand{\cdp}{\ma}
\newcommand{\cd}{\shortlist}
\newcommand{\dl}{L}
\newcommand{\vr}{r}

\begin{document}

%%% The following commands remove the headers in your paper. For final 
%%% papers, these will be inserted during the pagination process.

\pagestyle{fancy}
\fancyhead{}

%%% The next command prints the information defined in the preamble.

\maketitle 

%%%%%%%%%%%%%%%%%%%%%%%%%%%%%%%%%%%%%%%%%%%%%%%%%%%%%%%%%%%%%%%%%%%%%%%%

\section{Introduction}
Shortlisting is the process of selecting a subset of alternatives from a larger pool, for the purposes of further consideration or final decision-making. This process involves two groups of actors: {\em agents}, who will eventually vote over the options included in the shortlist to make the final selection, and {\em shortlisters}, who prepare the shortlist based on full or, more commonly, partial knowledge of agents' preferences, properties of the alternatives, and possibly their own preferences. The shortlisters may be representative of the agents' population,  
or they may be experts, or even AI systems.

\begin{ex} %[Shortlisting in participatory budgeting]
    \label{ex:realworld}
    Participatory Budgeting (PB) in Cambridge, MA~\citep{cambridgePB2025} receives over 1,000 project proposals in each cycle. During a four-month Proposal Development phase, volunteer shortlisters work with citizens and government officials to ``review, research, and develop submitted ideas to create final ballots.'' They narrow down the candidate set according to the need, impact, and feasibility of the proposals. Eventually, around 20  proposals are selected for a city-wide vote.
\end{ex}
Shortlisting is a general and commonly applied procedure in a wide range of social choice and multi-agent system scenarios. For example, in Reinforcement Learning with Human Feedback (RLHF), LLMs generate shortlists from a wide range of possible outputs to elicit preference from human workers. In a hiring process, applicants are screened on the basis of their track record before an interview. In school choice settings, students shortlist their target schools before a match is conducted.
\begin{figure*}
    \centering
\includegraphics[width=.8\textwidth]{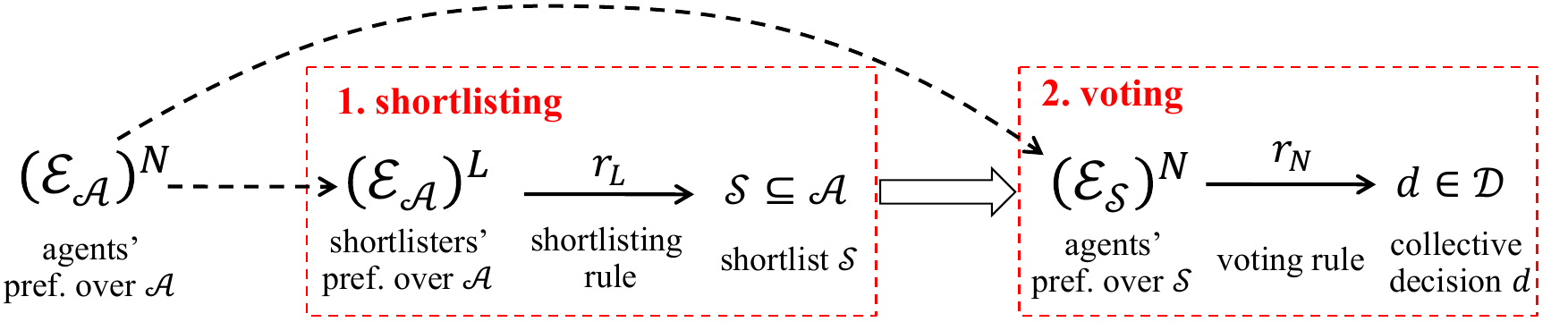}
\caption{Basic framework of shortlisting.\label{fig:shortlisting}}
\end{figure*}

The rise of participatory, generative, and AI-augmented collective decision-making~\citep{cambridgePB2025,fish2024generative,tessler2024ai}, and, in particular, 
direct democracy applications~\cite{helfer2021does,aitamurto2017value,wike2022global}, where the participants are often empowered to propose their own candidates, requires well thought-out shortlisting procedures. Indeed, democratic innovation may expand the candidate space, yet excessive choices impose heavy cognitive burdens on agents, potentially interfering with their ability to discover and communicate their true preferences~\citep {cunow2021less,augenblick2016ballot}. However, current practice relies on ad-hoc, 
informal, time-consuming and often non-transparent shortlisting methods~\citep{cambridgePB2025,oidp2020korea,seattle2024pb}. Consequently, there is an urgent call for a principled study of efficient and fair shortlisting procedures that aim to reduce cognitive burden, enable fair collective decisions, encourage broader participation, and ultimately build trust in democratic systems.

% We believe shortlisting is a particularly novel and important social choice problem. Modern participatory, generative, and AI-augmented collective decision making increasingly enriches the space of candidates~\citep{cambridgePB2025,fish2024generative,tessler2024ai}. However, overloading choices imposes a heavy cognitive burden on agents, distorting their true preferences~\citep{cunow2021less,augenblick2016ballot}. Such conflict creates an urgent need for a well-performed shortlisting procedure. 
% Efficient and fair shortlisting reduces cognitive burden, empowers efficient and fair collective decisions, encourages broader participation, and ultimately builds trust in democratic systems.

It is tempting to view shortlisting as an instance of {\em multiwinner voting}. This fields offers rich literature on both screening the most competitive candidates~\citep{gehrlein1985condorcet,aziz2017condorcet,barbera2008choose,bredereck2021coalitional,lackner2025approval} and electing fair and proportional committees~\citep {lackner2023multi,elkind2017properties,aziz2017justified,sanchez2017proportional}. However, we posit that shortlisting has  a fundamentally different goal: providing a good decision space (candidate sets) for subsequent voting is qualitatively different from directly determining the final set of winners. The following example illustrates this difference, by showing that desiderata suitable for multiwinner voting outcomes may be inappropriate in the context of shortlisting.

% \Qishen{shortlisting is a general problem not only in PB. }

% [One may immediately ask why this is no a multi-winner voting or run-off voting?]

% These real-world scenarios reveal the challenges in the shortlisting problem as well as the gap between existing research and practice. 

% \paragraph{vs. Multi-winner Voting} 
% One may immediately come up with the idea that shortlisting is a special type of multi-winner voting, which has become a heating topic in the past decade [cite]. 
% However, despite the similarity in formalities, the goal of shortlisting is fundamentally different from traditional multi-winner voting. Multi-winner voting aims to make a good final decision directly, with the outcome representing the collective decision of the agents. 
% Our shortlisting problem, on the other hand, aims to provide a good decision space (candidate) set for the subsequent voting. This difference leads to the different desiderata between the two scenarios. For example, the proportionality axiom, where agents get their proportional welfare, may not be suitable for the shortlisting, as a proportional shortlist does not necessarily lead to a proportional outcome. 
\begin{ex}
\label{ex:fail}
Consider a PB instance with five equally priced projects ($A$--$E$) and four agents, whose preferences are shown in   Table~\ref{tbl:intro}.

\begin{table}[htp]
\caption{Agents' preferences.\label{tbl:intro}}
\begin{tabular}{@{}ll@{}}
\toprule
Agent & Preferences \\ \midrule
1 & $A \succ E \succ$ others \\
2 & $B \succ E \succ$ others \\
3 & $C \succ E \succ$ others \\
4 & $D \succ E \succ$ others \\ \bottomrule
\end{tabular} 

\end{table}
Suppose all agents act as shortlisters, and the task is to shortlist four out of five projects, with the aim of selecting two projects to be funded in the second stage.  If we simply view shortlisting as multi-winner voting and choose a committee from the core~\citep{cheng2020group}, a well-accepted solution for multi-winner voting, then we must choose $\cd = \{A, B, C, D\}$.  However, no subset of two projects  from $\cd$ lies in the core w.r.t.~the agents' preferences over all alternatives, as any core committee must include $E$. % In contrast, $\{A, E\}$ is a feasible final outcome in the core, but it requires including $E$ in the shortlist.
\end{ex}

Another family of social choice methods that may appear to be  useful for shortlisting proceeds by delegating decisions to a subset of the agents:  
this includes representative democracy~\citep{abramowitz2025flexible,revel2024selecting} (representatives make decisions directly for the agents) and liquid democracy~\citep{miller1969program,brill2019interactive,golz2021fluid,kahng2021liquid,christoff2017binary} (agents delegate their voting powers to other agents). While these approaches reduce the cognitive burden on the agents (which is indeed the primary goal of shortlisting), they are problematic in that
representatives may have incentives to make selfish decisions in conflict with social welfare~\citep{strom2000delegation}, and delegation can lead to power concentration or even create a ``super agent'' that dominates the vote~\citep{golz2021fluid,kling2015voting};
moreover, limited knowledge of agents' full preferences hinders good decisions under both approaches. At the other end of spectrum is run-off voting~\citep{fishburn1981approval,freeman2014axiomatic,bartholdi1991single}, where agents shortlist by themselves; however, 
it is not more cognitively efficient, as the agents still need to consider the full candidate set when reporting their preferences. 

While there has been a large body of literature on agents directly making a collective decision~\citep{Brandt2016:Handbook}, to the best of our knowledge, there is little principled study on shortlisting, despite shortlisting being an important and multidisciplinary challenge with many real-world applications. An important exception here is the work of \citet{rey2021shortlisting}, who study shortlisting and voting in PB together as an end-to-end problem. However, their model assumes that the shortlisters have full knowledge of agent preferences. In contrast, we view shortlisting more broadly as a collective decision problem where access to full preferences is limited, which connects to work on distortion~\citep{procaccia2006distortion,anshelevich2018approximating,anshelevich2021distortion}, the preference bottleneck problem~\citep{Boutilier06:Constraint,Boutilier2002PODMP,Cramton06:Combinatorial}, and informed voting with incomplete information~\citep{Austen96:Information,Feddersen97:Voting,han2023wisdom}.  

We believe that research on shortlisting is both tractable (as it can build on many strands of existing work) and timely, given the growing popularity of participatory budgeting and other forms of direct participation. Crucially, we feel that the design of shortlisting procedures can benefit from both social choice ideas and concepts and modern machine learning techniques. Arguably, while voting bodies may be reluctant to adopt AI-inspired methods to aggregate the agents' preferences due to their desire for full transparency of the decision-making process, this may be less of an issue for the shortlisting stage; indeed, a well-documented shortlisting method that makes use of  machine learning tools may be {\em more} transparent than the status quo.
 % Compared to traditional social choice settings, shortlisting has less stringent requirements and more varied criteria across different scenarios. The intermediate nature of shortlisting also opens opportunities to leverage modern techniques---such as black-box predictions, large language models, and learning-based approaches---to bridge theoretical results into real-world practice.

This blue-sky paper aims to attract attention to the challenge of shortlisting, differentiate it from related problems, formulate initial intuitions, and, most importantly, serve as a call to arms. In what follows, we will discuss the problem setting (Section~\ref{sec:setting}), the desiderata (Section~\ref{sec:axiom}), the aggregation methods (Section~\ref{sec:aggregation}), and strategic aspects (Section~\ref{sec:incentives}). In each section, we will highlight the differences with prior work and provide some thoughts on challenges and future directions.

\section{The Shortlisting Framework}
\label{sec:setting}
\myparagraph{General social choice.} Let  $N = [n]   = \{1 ,2, \ldots, n\}$ denote the set of all agents. Let $\cdp$ denote the set of {\em alternatives} and $\mL (\cdp)$ denote the set of all full rankings over elements in $\cdp$. Let $\prefspace_\cdp$ denote the {\em preferences space} and $\decspace_\cdp$ denote the {\em (collective) decision space}. We suppress $\cdp$ from the notation when it is clear from the context. The classical social choice framework studies the design and analysis of a {\em voting rule} $r:{\prefspace}^n\ra \decspace$. For illustration purposes, this paper focuses on two types of preference space: {\em approval preference} ($\prefspace = \{0, 1\}^{\cdp}$) and {\em linear orders}  ($\prefspace =\mL(\ma)$). 

%A set $N = [n]$ of agents wants to make a collective decision from a {\em decision space} $\decspace$. 
\myparagraph{Basic Shortlisting Framework.} Figure~\ref{fig:shortlisting} illustrates the basic  shortlisting framework. The collective decision is made in two stages: the {\em shortlisting} stage and the {\em voting} stage. 
In the shortlisting stage, a group of {\em shortlisters}, denoted by $\dl$, use a shortlisting rule $\vr_L$ to select a {\em shortlist} of alternatives, denoted by  $\shortlist\subseteq \cdp$. The shortlisters may be sampled from $N$ (as representatives), chosen from outside (as experts), or a mixture. In the voting stage, the agents in $N$ use a voting rule $\vr_N:\prefspace_{\shortlist}^N\ra \decspace_{\shortlist}$ to make a collective decision. Notice that in the voting stage, the preference space and decision space are based on $\shortlist$ instead of all alternatives $\cdp$. We often assume that the shortlisters' preferences over $\cdp$ and agents' preferences over $\shortlist$ are influenced by the agents' preferences over $\shortlist$. Specifically,  each agent has preferences over $\cdp$ and converts it to preferences in $\prefspace_{\shortlist}$ in the voting stage. For example, under approval preferences, suppose an agent approves $A\subseteq \cdp$. Then in the voting stage, her (approval) preferences can naturally be $A\cap \shortlist$.

%The voting stage can represent a ranked-choice voting, a committee election, or participatory budgeting, etc. 

%For the purpose of illustration, we assume that agents and shortlisters share the same preferences space $\prefspace$. In this paper, we will focus on two types of preferences: approval preference ($\mV = \{0, 1\}^{\cdp}$) and ranking ($\mV =\mL(P)$). Note that practically, it is difficult for a agent to form a preference for all candidates in $P$. For a subset $\cd \in \cdp$, we define $\mV_{\cd}$ be the preference space on $\cd$. We use $\vt_\cdp$ and $\vt_\cd$ to denote agents' preferences on $\cdp$ and $\cd$, respectively, and $\vt_\cdp^\dl$ to denote the shortlisters' preferences on $\cdp$. For approval preference, we also use $\apr_i \subseteq \cdp$ to denote the set of candidates in $\cdp$ that agent $i$ approves. For a ranking $v_i$ and a candidate $j \in v_i$, let $\ell(v_i, j)$ be the ranking of candidate $j$ in $v_i$. We also use $j \succ_i j'$ to denote that agent $i$ prefers $j$ to $j'$ in $v_i$. 

% Each candidate has a cost, and let $\cost: \cdp \to \mathbb{R_+}$ be (additive) cost function. We may assume a unit cost for candidates when the scenario is not sensitive to costs.
% The voting rule $\vr_N$ takes the shortlist $\cd$, the cost function $\cost$, and the votes of all agents in $N$ (denoted as $V_C \in \mV_{\cd}^n$) and outputs a winner set $\wn$. In the rest of the paper, we will omit $c$ in the input of $r$ unless otherwise stated.  

\myparagraph{Extensions.} The basic shortlisting framework can be extended in various ways. For preference modeling, we can model shortlisters' and agents' preferences as probability distributions (for example, shortlisters' preferences $(\prefspace_{\cdp})^L$ may be drawn from a biased distribution with respect to agents' preferences $(\prefspace_{\cdp})^N$). We can also consider scenarios where the preference space and the voting space differ (for example, agents possess cardinal utilities but are required to vote with linear orders), connecting to work on sincerity~\citep{endriss2013sincerity,endriss2009preference} and distortion~\citep{anshelevich2021distortion}. For feasibility constraints, we may consider cardinality constraints, budget constraints, or external regulations that a shortlist must satisfy. For social welfare, we may define different welfare functions to evaluate the shortlist and the collective decision under different criteria. For additional information, we can incorporate information beyond shortlisters' preferences alone, such as black-box predictions, preference queries, or even LLM summarizations.

\section{Desiderata}
\label{sec:axiom}
Following the tradition of social choice theory, it is natural to approach the problem of shortlisting from an axiomatic perspective~\citep{Plott76:Axiomatic}, i.e.,  
by formulating desirable properties (axioms) of shortlisting procedures.
Thus, the central question of this section is:
\begin{center}
   \em  What are the desiderata for shortlisting? 
\end{center} 

Example~\ref{ex:fail} illustrates that axioms defined for multiwinner voting need not be suitable for shortlisting: indeed, such axioms impose constraints on the shortlist, but not on the collective decision.
Accordingly, we need to formulate desiderata that consider the entire decision-making process, as illustrated in Figure~\ref{fig:shortlisting}. Ideally, we would like the collective decision $d\in\decspace$ and the shortlist $\shortlist\subseteq\cdp$ to be good for the agents $N$ and for the shortlisters $\dl$. This intuition naturally allows us to systematically investigate four types of criteria that will be explored in the rest of this section. Each type depends on {\em what} is desirable for {\em whom}. %Each of the following subsections elaborates on one combination. 

%Different from traditional social choice problems, shortlisting has two voting principals (shortlisters and agents) and two phases of voting outcomes (the shortlist and winners). Therefore, we can categorize all the possible desiderata into four dimensions, which we discuss in detail in this section. 

%Another natural question is to study the compatibility between different axioms, which has been a common paradoxical phenomenon in other social choice models [cite]. 

%\begin{RQ}   Are the recovery-desiderata compatible with each other?\end{RQ}

% \begin{table}[htbp]
% \caption{Four dimensions of shortlisting desiderata\label{tbl:axiom}}

% \begin{tabular}{@{}lll@{}}
% \toprule
%  & Winner & Shortlist \\ \midrule
% agents & \textbf{Recovery} (Section~\ref{sec:recovery}) & Section\ref{sec:3_2} \\
% Shortlisters & Section~\ref{sec:3_3} & Section \\ \bottomrule
% \end{tabular}
% \end{table}

\subsection{Outcome: the Agents' Perspective}
\label{sec:recovery}
Perhaps the most important type of desiderata are the ones that evaluate the collective decision $d$ obtained at the end of process, from the perspective of the agents $N$. 
Specifically, given an axiom $X$, we consider a shortlist $\shortlist$ acceptable if the decision the agents in $N$ make when selecting from $\shortlist$ satisfies the axiom $X$ with respect to their full preference profile over $\cdp$. 
%Leveraging existing literature in social choice, naturally, we want the whole process to be as good, often measured by the satisfaction of an axiom $X$, as if all agents' preferences over $\cdp$ are completely known. This leads to the following class of axioms. %, which we call {\em recovery} axioms.
%\lirong{can we give names to other three types of axioms?}

%Among the four dimensions, we believe the most important is ``whether a shortlisting mechanism can help reach an efficient and fair final outcome (winner set) $\wn$ with respect to the preferences of the agents $\vt_\cdp$'', as it is closer to the fundamental goal. Given the gap between the large spectrum of existing desiderata in the social choice literature, where agents directly elect the outcome, and the intermediate nature of a shortlisting mechanism, we propose the following $X$-Recovery framework that translates existing social choice axioms into the shortlisting scenario. 

% The existing literature in social choice provides a large spectrum of desiderata covering different dimension in ``good'' decisions. To fit these desiderata into the shortlisting framework, we want the good properties to be in the winner set $\wn$ rather than the intermediate shortlist $\cd$ to guarantee a good final outcome. Therefore, we propose the following $X$-Recovery framework that translates existing social choice axioms into the shortlisting scenario. 

\begin{dfn}[$X$-Recovery]
\label{dfn:X-recovery}
    Given an axiom/desideratum $X$ defined based on agents' full preferences in $\prefspace_\ma$ and a voting rule $r_N$, % a shortlist $\shortlist$ satisfies $X$-recovery if $\wn = \vr(\cd, \vt_\cd)$ satisfies $X$ on the full candidate set $\cdp$ and full preference set $\vt_\cdp$. 
     a shortlisting procedure satisfies {\em $X$-recovery}  if for every collection of agents' full preferences $P_{\cdp}^N\in \prefspace_{\cdp}^N$, $X$ is satisfied by the collective decision $r_N(P_{\shortlist}^N)$, where $\shortlist = r_L(P_{\cdp}^L)$, and $P_{\cdp}^L$ and $P_{\shortlist}^N$ are induced by~$P_{\cdp}^N$.
\end{dfn}
While this definition is a bit vague, it follows the intuition in Example~\ref{ex:fail} that a ``good'' shortlisting procedure should provide a decision space so that a ``good'' decision can be made as if agents' full preferences are known. It allows us to directly leverage existing axioms to evaluate a shortlisting mechanism.  Likewise, following a similar idea, we can define a recovery axiom associated with a voting rule $\vr$, which requires the shortlist to recover the outcome as if all agents directly vote on all alternatives under voting rule $\vr$. We omit the definition due to space constraints. 

%\begin{ex}We follow the setting in Example~\ref{ex:fail} to illustrate our recovery desiderata. Suppose the shortlist $\cd = \{a, c, d, e\}$. $\cd$ satisfies weak core-recovery, as it contains $\{a, e\}$. When the voting rule $r$ is 2-plurality, the winner $W$ will be a subset of $\{a, c, d\}$ and not be in the core. In this case, $\cd$ does not satisfy core-recovery. On the other hand, when the voting rule is 2-Borda, $e$ will be elected as a winner, and $\cd$ satisfies core-recovery. \end{ex}

Intuitively, a larger shortlist should make it easier to satisfy a recovery desideratum. On the other hand, larger shortlists place a cognitive burden on the agents and sacrifice efficiency. We will discuss this tradeoff in Section~\ref{sec:tradeoff}.

\subsection{Shortlist: the Agents' Perspective} 
\label{sec:3_2}
While recovery desiderata are important, we believe that they do not fully exhaust what it means for a shortlisting procedure to be acceptable. Example~\ref{ex:axiom2} motivates the need for the shortlist itself to be attractive with respect to the agents' preferences.

\begin{ex}
    \label{ex:axiom2}
    Consider a PB scenario with 20 projects where a shortlist of 5 candidates is first created, and then agents vote to select a single candidate by means of approval voting, i.e., so that a candidate with the largest number of approvals is selected. Suppose 50\% of the agents approve project $A$, and another 30\% of the agents approve project $B$. All other projects receive very few approvals. A shortlist $\cd$ that contains $A$ but not $B$ leads to the same collective decision as if all agents vote directly. However, intuitively, $\cd$ is not a good shortlist, as supporters of $B$ are likely to feel underrepresented, manipulated, and discouraged, and may abstain from participating in the future PB cycles.
\end{ex}

Various efficiency, fairness, and diversity axioms can be investigated, leveraging the large literature on single-winner voting (such as Pareto efficiency), multi-winner voting (such as proportionality and justified representation~\citep{aziz2017justified,sanchez2017proportional,aziz2017condorcet,elkind2024price}), and matching problems (such as stability under constraints~\citep{Gale62:College}). 
%EE removed fair division, as we are in a public goods setting
%and fair division (such as envy-freeness and its relaxations~\citep{budish2011combinatorial,caragiannis2019unreasonable,hosseini2020fair}).  %We enumerate some axioms for the shortlist $\cd$ under different dimensions. The first axiom is Pareto Optimality, which requires that no candidate in the shortlist is unanimously preferred by a candidate outside. This axiom follows the idea that the shortlist should be efficient and present the best candidates to the agents. 

\subsection{Outcome: the Shortlisters' Perspective}
\label{sec:3_3}
The reason to look into axioms on the final outcome with respect to shortlisters' preferences is two-fold. First, shortlisters have incentives to influence the outcome via the shortlist. We hope that the strategic behavior of shortlisters can benefit rather than harm the final decisions; this will be discussed in more detail in Section 5. Second, the preferences of shortlisters can be viewed as an accessible approximation of those of agents. Therefore, guaranteeing a good outcome for shortlisters may be a practical way to reach a reasonably good outcome with respect to the agents' preferences. 

%For ``winner + shortlister'' axioms, we can also leverage existing axioms and define a similar framework to recovery by changing the agents into shortlisters. Besides, we are particularly interested in axiom related to strategic behaviors. For example, the following axiom requires no shortlisters can benefit from misreporting in the shortlisting stage. 

%\begin{dfn}[Incentive Compatible]  A shortlist mechanism $\mM$ is incentive compatible if for any shortlisting instance and any shortlister $i \in \dl$, $i$ prefers the outcome when he/she vote truthfully $\vr(\cd = \mM(\cdp, \vt_\cdp^\dl), \vt_\cd)$ over that when he/she take any other action $\vr(\cd' = \mM(\cdp, (v_i, \vt_{\cdp, -i}^{\dl})), \vt_{\cd'})$, where $\vt_{\cdp, -i}^{\dl}$ is the votes of other shortlisters. \end{dfn}

\subsection{Shortlist: the  Shortlisters' Perspective}
\label{sec:3_4}
This combination is similar to multi-winner voting, so researchers can start with axioms for multi-winner voting. One caveat is that for shortlisting, the size of the shortlist is variable. Additionally, we may also consider learning-augmented shortlisting, as discussed later in Section~\ref{sec:aggregation}.

\section{Aggregation}
\label{sec:aggregation}
Having defined the desiderata, the next natural question is 
\begin{center}
    \em Can we identify shortlisting procedures that satisfy our axioms?
\end{center}
As in classical social choice, we start by considering non-strategic agents and shortlisters; we defer discussion of strategic aspects to the next section.  

The design and analysis of the shortlisting rule $r_{L}$ and the voting rule $r_N$ in the shortlisting framework (Figure~\ref{fig:shortlisting}) can be built upon the vast literature on social choice, such as multi-winner rules~\citep{lackner2023multi} (e.g., approval voting and method of equal shares~\citep{peters2020proportionality}), single-winner voting, matching, and recommender systems, just to name a few. We foresee a rich and fruitful research agenda on evaluating the satisfaction of existing rules w.r.t.~existing axioms and their recovery variants (Section~\ref{sec:axiom}). Such studies can be based on  worst-case analysis as in classical social choice theory~\citep{Plott76:Axiomatic}, average-case analysis under popular distributions over preferences~\citep{Diss2021:Evaluating,boehmer2023properties}, or smoothed analysis~\citep{xia2020smoothed,flanigan2023smoothed}. The design of novel shortlisting processes to satisfy these axioms is another promising direction for future work, for which the automated mechanism design paradigm can be helpful~\citep{Conitzer03:Automated,Mohsin2022:Learning,Armstrong2019:Machine,lanctot2025soft}.

Additionally, below we highlight two specifically interesting, and somewhat unique aspects of shortlisting: cognitive efficiency and shortlisting with predictions.

\subsection{Cognitive Efficiency}
\label{sec:tradeoff}
As mentioned earlier, determining the size of the shortlist is a trade-off between desiderata and cognitive efficiency. A larger shortlist provides a larger decision space for the agents, and makes it easier to satisfy the recovery desiderata. On the other hand, a larger shortlist increases the cognitive burden on the agents, as they need to form opinions on a larger set of candidates. The following example illustrates these tradeoffs and connects two extremes of shortlist size to two traditional social choice problems. 

\begin{ex}
    Consider a shortlisting scenario where a shortlist with $k$ candidates is first created, and $k'$ winners are then elected. When $k' = k$, the shortlisters make the decision for the agents; this is what happens in representative democracy. On the other hand, when $k = |\cdp|$, the shortlisters perform no shortlisting, and the scenario is a direct vote on all candidates. Therefore, $|\cdp| \ge k' \ge k$ spans a spectrum of different shortlisting scenarios from direct democracy to representative democracy.
\end{ex}
We believe that analyzing the tradeoff between cognitive efficiency and other desiderata is a key challenge in the design of shortlisting procedures, as done in seminal works on college admission~\citep{chen2017chinese,chen2019chinese}.

\subsection{Shortlisting with Predictions} 
Algorithms with predictions~\citep{mitzenmacher2022algorithms} is an emerging paradigm for the design an analysis of online algorithms, partly motivated by the availability of efficient machine learning tools. Briefly, algorithms with predictions leverage potentially imperfect predictions about the input to achieve better performance, while providing guarantees even if the predictions turn out to be incorrect. Despite its popularity and success in many subfields, surprisingly few papers apply this paradigm to problems in social choice (see, however, \citep{filos2025utilitarian,berger2024learning}), perhaps because preference aggregation is usually viewed as an offline problem. However, shortlisting exhibits temporal structure that makes it suitable for this approach. Indeed, the shortlisters may proceed  by predicting voters' preferences, and, in some cases, the features of the candidates themselves: e.g., in participatory budgeting setting the initial proposals do not come with precise costings, so one may try to predict the project costs.  

More generally, there is a wide range of predictions one may consider for shortlisting, including (distribution of) agents' full preferences over $\prefspace_{\cdp}$, the final collective decision based on agents' full preferences, or cost of a project for participatory budgeting. With the help of predictions, a shortlisting mechanism may circumvent impossibility theorems of traditional mechanisms and achieve stronger axiomatic properties. However, given its black-box nature, the prediction has no accuracy guarantee, and blindly following the prediction may lead to an arbitrarily bad outcome~\citep{purohit2018improving,lykouris2021competitive}.  Following previous work~\citep{purohit2018improving,lykouris2021competitive,xu2022mechanism,agrawal2022learning,balkanski2024online}, the goal is to design shortlist mechanisms that have high performance under accurate prediction (consistency) and reasonable performance under arbitrary prediction (robustness).

\section{Incentives}
\label{sec:incentives}

Incentives and strategic behaviors are another important aspect in the principled study of shortlisting. The central question is:

\begin{center}
    {\em How can we handle  strategic behaviors of agents and shortlisters in the design and analysis of shortlisting?}
\end{center}

\myparagraph{Model the incentives.} The incentives of agents and shortlisters can come from many sources. For example, a shortlister may be altruistic and care about the social welfare or loyal to the agents he/she represents, an AI agent aims to optimize a loss function, and an expert shortlister prioritizes the quality of the candidates, just to name a few. Importantly, even if a shortlister is altruistic, they need not be truthful, especially if they know that their own preferences may be quite different from the voters' preferences. 

\myparagraph{Game-theoretic analysis.} Under a game-theoretic lens, we model the shortlisting process as a game, analyzing the strategic behavior of shortlisters, agents, or both. For example, one can form a two-stage extensive-form game to analyze the strategic interactions between shortlisters and agents, or form a Bayesian game to characterize shortlister behaviors when they receive distributional information regarding the agents' preferences. Possible research includes determining equilibria, calculating the incentive ratio~\citep{chen2012incentive}, or designing incentive-compatible mechanisms.

\myparagraph{The effect of strategic behavior.} Another natural question is how strategic behaviors affect reaching good collective decisions. While a large literature focuses on the negative effects of strategic behavior, such as the price of anarchy~\citep{Branzei13:How,bailey2025price} and the complexity of manipulation~\citep{Bartholdi92:How,Bartholdi89:Computational}, strategic behavior can also be beneficial to reach a good outcome, as it is illustrated in voting with incomplete information ~\citep{han2023wisdom} and iterative voting~\citep{kavner2021strategic}. For example, one interesting direction exclusively for shortlisting is that, given agents being strategic, do strategic shortlisters help or harm in making good collective decisions compared to non-strategic ones? Understanding these interactions is key to designing shortlisting mechanisms that remain robust and beneficial even when participants act strategically.

\section{Summary and Future Vision}
In this paper, we take a first step toward a principled study of shortlisting as an important yet underexplored problem in social choice and multi-agent systems. We propose a unified model and examine the key aspects of desiderata, aggregation rules, and strategic incentives in the shortlisting problem. Many important topics remain open for future investigation, including the explainability and accountability of shortlisting mechanisms, applications to apportionment, fair division, and other social choice settings, as well as integration with modern techniques such as recommender systems and large language models. On the practical front, how to implement new shortlisting rules and how to and educate the policy makers and stakeholders about shortlisting, or more generally,  social choice and democracy, are important tasks. Our goal in writing this paper was to inspire future work on establishing a solid foundation for designing shortlisting mechanisms that are fair, transparent, and grounded in principled theory—not only for political systems, but for the wide range of collective decision-making scenarios emerging in organizations, online platforms, and AI-assisted governance.

%\Qishen{Leave blank at this moment. Will discuss about a broad social choice function, such as portioning, ranking, matching, fair division... }

%\lirong{We need a quick summary and strong finish, and mention that we didn't discuss other important topics such as explainability etc.}

\bibliographystyle{ACM-Reference-Format}
\bibliography{references,newref}
\end{document}